%% file: ICIP_main2.tex
\documentclass{article}
\include{macros}

\usepackage{spconf}
\usepackage{subcaption}
\usepackage{algorithm}
\usepackage{graphicx}  
\usepackage{color}
\usepackage{xcolor}
\usepackage{tcolorbox}
\usepackage{comment}
\usepackage{amsmath}
\usepackage{array}
\usepackage{multirow}
\usepackage{tabularx}
\usepackage{xcolor}
\usepackage{caption}
\usepackage{subcaption}
\usepackage{times}
\usepackage{latexsym}
\usepackage{pgfplots}
\usepackage{pgf-pie}
\usepackage[T1]{fontenc}
\usepackage[utf8]{inputenc}
\usepackage{tikz}
\usetikzlibrary{positioning}
\usepackage{microtype}
\usepackage{pgf-pie}
\usepackage{inconsolata}
\usepackage{graphicx}
\usetikzlibrary{decorations.pathreplacing}
\usepackage[justification=centering]{caption}
\definecolor{darkgreen}{rgb}{0.0, 0.5, 0.0}

\newcommand{\arrowUpDarkGreen}{\textcolor{darkgreen}{$\uparrow$}}
\newcommand{\arrowDownDarkGreen}{\textcolor{darkgreen}{$\downarrow$}}

\definecolor{customNavy}{RGB}{0, 0, 128}   
\definecolor{customTeal}{RGB}{0, 128, 128} 

\definecolor{mygr}{rgb}{0.6,0.4,0.0}
\definecolor{my1color}{rgb}{0.6,0.4,0.0}
\definecolor{mycolor1}{rgb}{0.00000,0.44700,0.74100}%
\definecolor{mycolor2}{rgb}{0.85000,0.32500,0.09800}%
\definecolor{mycolor3}{rgb}{0.45000,0.62500,0.19800}%
\definecolor{mycolor4}{rgb}{0.75000,0.1500,0.100}%
\definecolor{RYB1}{RGB}{218,232,252}
\definecolor{RYB2}{RGB}{245,245,245}
\definecolor{RYB3}{RGB}{145,200,100}
\definecolor{RYB4}{RGB}{108,142,191}

\title{Can Large Language Models Challenge CNNs in Medical Image Analysis?}
\threeauthors
 {Shibbir Ahmed}
	{Texas State University\\
	San Marcos, TX, USA}
{\hspace{-1cm}Shahnewaz Karim Sakib}
	{\hspace{-1cm}University of Tennessee at Chattanooga\\
	\hspace{-1cm}Chattanooga, TN, USA} 
 {\hspace{-1cm}Anindya Bijoy Das}
	{\hspace{-1cm}The University of Akron\\
	\hspace{-1cm}Akron, OH, USA} 
    
\begin{document}
%
\maketitle
\tikzset{
block/.style    = {draw, thick, rectangle, minimum height = 2em, minimum width = 2em},
sum/.style      = {draw, circle, node distance = 1cm},
sum1/.style      = {draw, circle, minimum size = 1.1 cm},
input/.style    = {coordinate},
output/.style   = {coordinate},
}
\input{abstract.tex}


\input{introduction2.tex}

\input{problem2}
\input{methodology2}

\input{results2}
\input{conclusion2}


%



\bibliographystyle{IEEEbib}
\bibliography{strings,refs}
\balance
\end{document}

%% file: macros.tex
\usepackage{xurl}
\usepackage{multirow}
\usepackage{caption} 
\usepackage{booktabs} 
\usepackage{amsmath}
\usepackage{graphicx, subfigure}
\usepackage[belowskip=-12pt]{caption}
\usepackage{rotating}
\usepackage{rotating}
\usepackage{tabularx}

\usepackage{wrapfig}
\usepackage{listings,chngcntr}
\usepackage{color, colortbl}
\usepackage{balance} 
\usepackage{lscape}
\usepackage{algorithm}
\usepackage[noend]{algpseudocode}
\algnewcommand\algorithmicforeach{\textbf{for each}}
\algdef{S}[FOR]{ForEach}[1]{\algorithmicforeach\ #1\ \algorithmicdo}

\usepackage{xspace}
\usepackage{framed}
\usepackage{syntax}
\usepackage[tikz]{bclogo}
\usepackage{soul}
\usepackage{flushend}
\usepackage{pifont}
\newcounter{NumObservations}
\stepcounter{NumObservations}
\definecolor{shadecolor}{rgb}{.9,.9,.9}

\usepackage{titlecaps}

\usepackage{tikz}
\usetikzlibrary{positioning}
\usetikzlibrary{shapes.geometric}
\usetikzlibrary{shapes.misc}

\usepackage{listings}
\usepackage{xcolor}

\definecolor{codegreen}{rgb}{0,0.6,0}
\definecolor{codegray}{rgb}{0.5,0.5,0.5}
\definecolor{codepurple}{rgb}{0.58,0,0.82}
\definecolor{backcolour}{rgb}{0.97,0.97,0.98}

\lstdefinestyle{mystyle}{
    backgroundcolor=\color{backcolour},   
    commentstyle=\color{codegreen},
    keywordstyle=\color{magenta},
    numberstyle=\tiny\color{codegray},
    stringstyle=\color{codepurple},
    basicstyle=\ttfamily\footnotesize,
    breakatwhitespace=false,         
    breaklines=true,                 
    captionpos=b,                    
    keepspaces=true,                 
    numbers=left,                    
    numbersep=5pt,                  
    showspaces=false,                
    showstringspaces=false,
    showtabs=false,                  
    tabsize=2
}

\lstdefinelanguage{Lambda}{%
  morekeywords={%
    if,then,else,fix 
  },%
  morekeywords={[2]int},   
  otherkeywords={:}, 
  literate={
    {->}{{$\to$}}{2}
    {lambda}{{$\lambda$}}{1}
  },
  basicstyle=\fontsize{9}{9}\ttfamily,
  keywordstyle={[2]},
  	breakatwhitespace=true,         
	keepspaces=true,                 
 	showspaces=false,                
 	showstringspaces=false,
  keepspaces,
  mathescape 
}[keywords,comments,strings]%

\usepackage{mathpartir}


%% file: abstract.tex
\begin{abstract}
This study presents a multimodal AI framework designed for precisely classifying medical diagnostic images. Utilizing publicly available datasets, the proposed system compares the strengths of convolutional neural networks (CNNs) and different large language models (LLMs).
This in-depth comparative analysis highlights key differences in diagnostic performance, execution efficiency, and environmental impacts.
Model evaluation was based on accuracy, F1-score, average execution time, average energy consumption, and estimated $CO_2$ emission.
The findings indicate that although CNN-based models can outperform various multimodal techniques that incorporate both images and contextual information, applying additional filtering on top of LLMs can lead to substantial performance gains. These findings highlight the transformative potential of multimodal AI systems to enhance the reliability, efficiency, and scalability of medical diagnostics in clinical settings.
\end{abstract}
\vspace{0.1 in}

%% file: introduction2.tex
\vspace{-0.15 in}
\section{Introduction}
\label{intro}
\vspace{-0.1 in}

The rapid evolution of artificial intelligence (AI) has ushered in transformative approaches for medical diagnostics, particularly in the analysis of diagnostic imaging data \cite{9363915}. Over the past decade, machine learning techniques have been applied successfully to tasks such as image classification, segmentation, and anomaly detection in modalities including X‐ray \cite{9376703}, computed tomography (CT) \cite{9420107}, and magnetic resonance imaging (MRI) \cite{9779760}. Early breakthroughs using convolutional neural networks (CNNs) \cite{9376703}, recurrent neural networks (RNNs), and autoencoders demonstrated the potential of deep architectures to learn hierarchical representations directly from raw images. 
Recent advancements in deep learning have introduced alternative architectures such as transformers \cite{hatamizadeh2022unetr}, capsule networks \cite{9514545}, and diffusion models \cite{10167641}, significantly enhancing the capabilities of medical image analysis. These innovative models excel in capturing complex patterns and dependencies in medical data, which is crucial for accurate diagnosis and effective treatment planning. Their successful application across various clinical scenarios demonstrates their potential in improving the precision and efficiency of healthcare interventions.

In parallel with these imaging advances, large language models (LLMs) \cite{thirunavukarasu2023large} have emerged as powerful tools for processing and interpreting clinical narratives, radiology reports, and other unstructured data sources \cite{thirunavukarasu2023large, hager2024evaluation, das2025hallucinations, kung2023performance}. LLMs complement image-based predictions to boost diagnostic accuracy by extracting and contextualizing patient information. For example, integrating LLMs with imaging architectures enables automated report generation and decision support, i.e., cross-referencing clinical findings with current literature, to reduce clinician workload and diagnostic errors \cite{hager2024evaluation}. Furthermore, the synergy between LLMs and advanced machine learning techniques enhances AI interpretability and explainability \cite{zhao2024explainability}, improving trust among healthcare professionals and enabling personalized treatments. Incorporating LLMs augments imaging model performance, streamlines diagnosis, and improves clinical workflows, ultimately enhancing patient care.

\begin{figure*}[t]
    \centering
    \subfigure[Chest X-ray]{\label{figoriginaldataset}
        \includegraphics[width=0.29\textwidth]{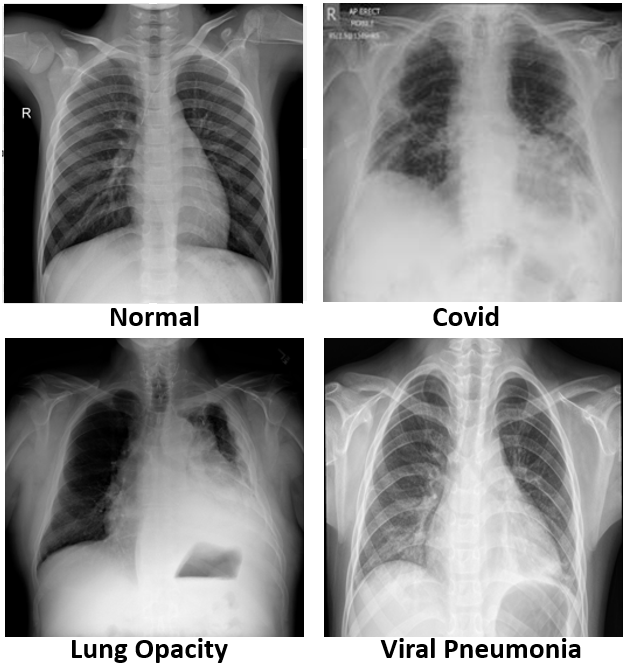}}
    \subfigure[Chest CT scan]{\label{fig:vfl1}
        \includegraphics[width=0.36\textwidth]{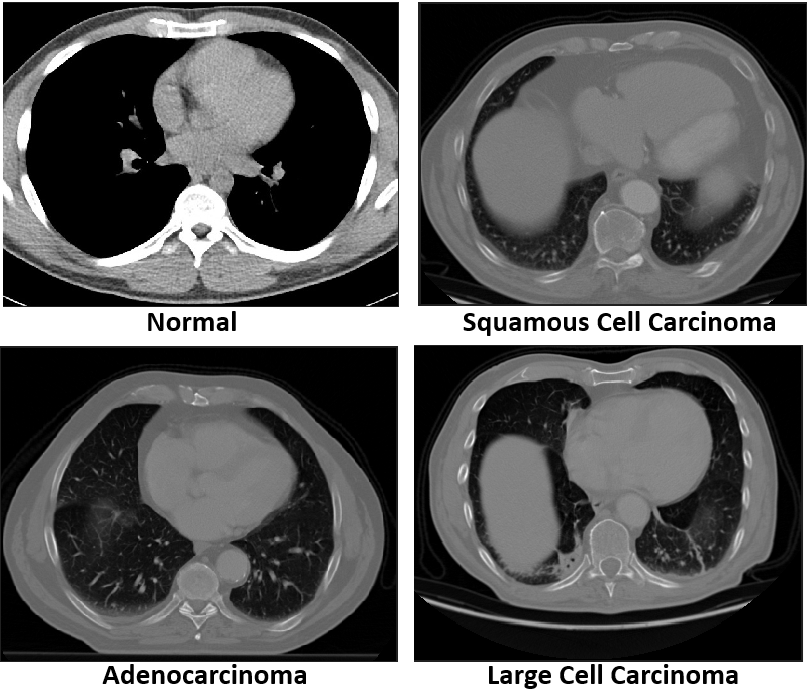}}
    \subfigure[Brain MRI]{\label{fig:vfl2}
        \includegraphics[width=0.26\textwidth]{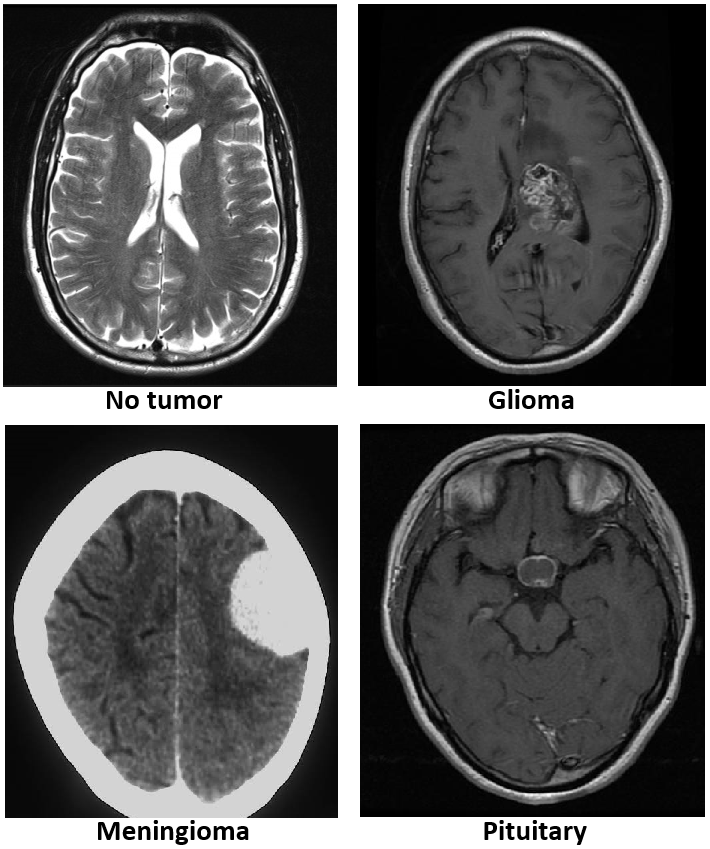}}
        \vspace{-0.1 in}
    \caption{\small Different Medical Imaging Modalities and Pathologies: 
    (a) Chest X-rays depicting normal lung anatomy, COVID-19 infection, lung opacity, and viral pneumonia. 
    (b) Chest CT scans for normal lung, squamous cell carcinoma, adenocarcinoma, and large cell carcinoma. 
    (c) Brain MRI scans highlighting conditions such as no tumor, glioma, meningioma, and pituitary abnormalities. 
    These exemplify typical radiological features used for training and evaluating AI-driven diagnostic models in medical imaging.}
    \label{figure1}
        \vspace{-0.1 in}
\end{figure*}

In this study, we investigate the reliability and performance of multimodal AI approaches for diagnostic image classification by comparing the effectiveness of CNN-based architectures with that of several benchmark large language models \cite{lee2025cxr, 10720422}, including OpenAI's GPT-4o, and Meta's LLaMA. We employ publicly available datasets comprising X-ray, CT, and MRI modalities (as shown in Fig. \ref{figure1}), which are meticulously partitioned into training, validation, and unseen testing sets. In addition, we explore these multiple LLMs to assess their potential in diagnostics, where few studies \cite{10522762} have concurrently addressed both utility and reliability.

Our systematic evaluation employs performance metrics such as accuracy, F1-score,  average execution time, average energy consumption, and estimated $CO_2$ emission to provide detailed insights into the strengths and limitations of CNNs and LLMs in real-world clinical scenarios. Additionally, we explore confidence scores as a measure of reliability, offering a comparative analysis of the trustworthiness of these models in diagnostic decision-making. 
This study contributes by:
\begin{itemize}
    \item Benchmarking CNNs and LLMs on classifying diverse medical imaging datasets, including image analysis by the LLM models for future study. \vspace{-0.03in}
    \item Analyzing computational efficiency, including runtime, energy consumption, and associated CO\textsubscript{2} emissions. \vspace{-0.03in}
    \item Highlighting trade-offs between model accuracy and resource demands, emphasizing considerations critical for practical and sustainable AI adoption in healthcare. \vspace{-0.03in}
\end{itemize}

In summary, our research findings provide useful benchmarks for researchers and valuable insights for medical professionals seeking to adopt AI-driven diagnosis, ultimately improving clinical workflow reliability and efficiency.

%% file: methodology2.tex
\section{Datasets and Related Works}
\label{approach}
In this section, we detail the datasets utilized in this study and review existing literature relevant to the classification of medical images.


\subsection{Dataset Selection}
To ensure a comprehensive evaluation, we select the following publicly available datasets covering a range of imaging modalities and diagnostic tasks. These datasets were curated to maintain class balance, diversity, and representativeness. 

\vspace{0.05 in}
\noindent \textbf{Chest X-Ray Dataset:} The COVID-19 Radiography Database \cite{chowdhury2020can} contains X-ray images labeled as COVID-19, pneumonia, lung opacity, and normal. This dataset is commonly used to benchmark AI models for classifying chest X-rays. \cite{rahman2021exploring}.

\vspace{0.05 in}
\noindent \textbf{Brain Tumor MRI Dataset:} The Brain Tumor MRI Dataset \cite{mridataset},  available on Kaggle, includes T1-weighted MRI images labeled as glioma, meningioma, pituitary tumor, or no tumor. This dataset enables the evaluation of models in brain tumor classification tasks.

\vspace{0.05 in}
\noindent \textbf{Chest CT Scan Dataset:} This dataset \cite{chestctscan}, also available on Kaggle, includes chest CT scans labeled as normal, adenocarcinoma, large cell carcinoma and squamous cell carcinoma. This enables us to work on different chest cancer detection.

Sample images are illustrated in Fig.~\ref{figure1}, which demonstrates the distinctions between different types of medical imaging: chest X-ray in Fig. \ref{figure1}(a), chest CT scan in Fig. \ref{figure1}(b) and brain MRI in Fig. \ref{figure1}(c).

\subsection{Related Works}
Next, we explore relevant studies focused on the analysis and classification of medical images, such as X-rays, CT scans, and MRI data, utilizing various LLM-based methodologies.

\vspace{0.05 in}
\noindent {\bf Transfer Learning and Pre-trained Models}: Recent studies have adapted pre-trained language models, like BERT and GPT, to extract and contextualize features from medical images. For example, \cite{khare2021mmbert} developed MMBERT explores on two visual question answering (VQA) datasets for radiology images, outperforms different previous ensemble models and provides attention maps which help in model interpretability.

\vspace{0.05 in}
\noindent {\bf Multi-modal Learning Approaches}: Multi-modal learning techniques have shown great promise in enhancing the analysis of medical images by integrating models trained on different types of data \cite{10288259}. For instance, incorporating neural networks with LLMs can enable a synergistic approach where the model benefits from the spatial recognition capabilities of neural networks while leveraging the contextual processing strengths of LLMs. The work in \cite{10614428} demonstrates the effectiveness of large visual language models in analyzing biomedical images such as brain MRIs, microscopic images of blood cells, and chest X-rays.

\vspace{0.05 in}
\noindent {\bf Fine-tuning on the Datasets}: Tailoring LLMs to specific medical tasks has proven to be an effective strategy \cite{10614428}. Fine-tuning models on specialized datasets, such as radiology reports, enhances their ability to accurately classify medical images, including CT scans of lung nodules.

\vspace{0.05 in}
\noindent {\bf Explainability and Ethical Considerations}: With AI’s growing role in healthcare, the demand for explainable AI models has surged \cite{nazi2024large}. Explanations generated by LLMs can help radiologists understand the AI’s reasoning process, potentially increasing trust and reliability in AI-assisted diagnostics \cite{zhao2024explainability}.

\section{Analysis of the Models}
\label{evaluation}
In this section, we present the various models employed in this study for classifying medical images, followed by an analysis of the corresponding classification processes.

\vspace{-0.1 in}
\subsection{Classification Approaches} In this study, we investigate two complementary paradigms for medical image classification: large language models (LLMs) and deep neural networks (DNNs). Each approach is tailored to address the unique challenges of medical imaging and is described in detail below.

\vspace{0.03 in}
\noindent \textbf{Large Language Models (LLMs):}
Traditionally employed for natural language processing tasks, LLMs have recently shown promise in multimodal applications, including image classification. By leveraging language-based embeddings, images can be transformed into descriptive representations that LLMs process to generate classification predictions. In this study, we explore methods to integrate image features directly into LLM architectures, focusing on adapting state-of-the-art models such as GPT-4o and Llama3.2-vision for medical diagnostics. We assess LLM feasibility in tasks typically dominated by image-centric models, identifying strengths and limitations in handling visual medical data.

\vspace{0.03 in}
\noindent \textbf{Deep Neural Networks (DNNs):}
DNNs, particularly convolutional neural networks (CNNs), have established themselves as a gold standard for image-based tasks, including medical diagnostics. For this study, we use state-of-the-art CNN architectures that have demonstrated high performance in medical imaging, such as the COVID-Roentgen CNN for chest X-rays and a TensorFlow-based CNN for brain tumor classification. 

Specifically, for the chest X-Ray dataset, we employed the COVID CNN model and optimized to classify COVID-19 and normal chest X-rays. Next, for brain tumor MRI classification, we used a TensorFlow-based CNN model capable of distinguishing benign (normal) and three different types of malignant (cancerous) growths in high-resolution images. For the chest CT scans, we utilized a fine-tuned CNN trained on the dataset to differentiate normal and three different chest cancer scans. All these DNNs are fine-tuned on the respective datasets to ensure optimal performance, emphasizing their ability to learn domain-specific features from relatively small datasets. Note that each dataset was split into training (80\%), validation (20\%), and unknown (20\%) subsets, following standard practices for medical image analysis. 


\subsection{Enhanced Data Filtering for LLM Classifications}

\input{filtering}

%% file: filtering.tex
\vspace{-0.1 in}
\label{sec:filtering}
Now we refine the data filtering process for LLM classification by implementing a multi-stage strategy that carefully screens and integrates contextual and feature details. This approach ensures only essential data (or features) is selected for analysis, thereby improving both the precision and consistency of classification outcomes. The enhanced methodology is outlined through the following step-by-step workflow:
\begin{enumerate}
    \item \textbf{Initial Label-Based Filtering:} We begin by enumerating the responses across the entire training set. From this dataset, we first extract only those samples corresponding to the desired label (e.g., ``normal'' in the context of COVID classification).
    \vspace{-0.05 in}

    \item \textbf{Confidence Thresholding:} To further refine the dataset, we associate each sample with a confidence score (such as the output of a softmax layer reflecting the model’s certainty in its prediction). We then discard samples with scores below a chosen threshold (e.g., $0.8$), thereby retaining only high-confidence, label-consistent samples that are more likely to be correctly classified and less prone to noise or ambiguity.
\vspace{-0.05 in}

    \item \textbf{Context and Feature Extraction:} Next, we examine these high-quality samples to identify key contextual elements and features (e.g., specific words, phrases, or structural patterns) that enable the Large Language Model (LLM) to classify them as the chosen label.
\vspace{-0.05 in}

    \item \textbf{Context Aggregation and Question Formulation:} We aggregate these contextual insights from all filtered samples and present them to GPT-4o. This step produces a consolidated understanding of the features driving classification decisions, as well as a set of targeted questions whose answers are critical for label determination.
\vspace{-0.05 in}

    \item \textbf{Application During Testing:} Finally, in the testing phase, we incorporate these targeted questions into the user prompt, rather than attempting to enumerate all possible features. By prompting the LLM to respond to the key questions directly, we streamline the classification process and increase the reliability of determining whether a given image matches the target label.
\end{enumerate}

\begin{figure}[!h]
\centering
	\includegraphics[scale = 0.42,trim={1cm 7.2cm 7.5cm 0.5cm},clip]{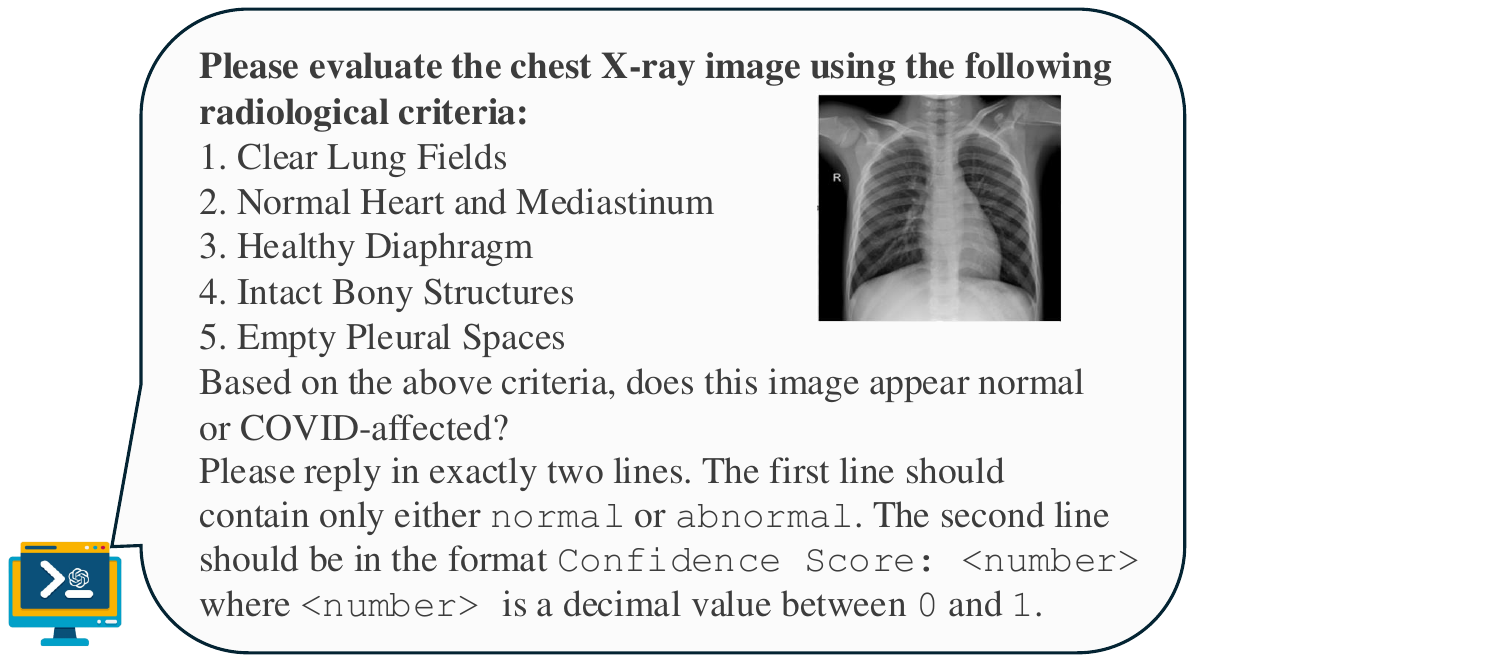}
	\footnotesize
    \vspace{-0.3 in}
	\caption{Context Aggregation and Question Formulation} 
    \vspace{0.1 in}
	\label{fig:chart2}
\end{figure}

\vspace{-0.1 in}
A sample user prompt is provided in Fig. \ref{fig:chart2}. Note that we also include a confidence score for each test image, thereby quantifying the model’s reliability in its predictions.



\vspace{-0.15 in}
This method streamlines the decision-making process by equipping the Large Language Model (LLM) with specific contextual information necessary for the classification task. Consequently, this targeted data provision not only accelerates the execution time but also lowers the energy required, enhancing the model's efficiency and environmental footprint.

%% file: results2.tex
\vspace{-0.15 in}
\section{Results and Discussions}
\vspace{-0.05 in}
\subsection{Evaluation Metrics}
The performance of CNNs and LLMs was evaluated using several key metrics. Accuracy is measured by the percentage of correctly classified samples, providing an overall performance indicator. Precision assessed the proportion of true positives among predicted positives, minimizing false positives, while Recall measured the proportion of true positives among actual positives, emphasizing the model's ability to identify relevant cases. The F1-Score \cite{10825564}, as the harmonic mean of precision and recall, offered a balanced measure of performance. Additionally, the Confidence Score \cite{guo2017calibration} analyzed prediction reliability using calibration curves. Finally, resource consumption, such as inference time and energy usage, was evaluated comprehensively to assess the models' efficiency and practicality in real-world healthcare applications.
\vspace{-0.1 in}

\label{sec:results}

\begin{table}[t]
\small
\centering
\caption{\small Performance comparison of LLM and CNN models in terms of accuracy, F-1 score and average confidence score (CS) across different medical imaging modalities}
\vspace{0.1 in}
\label{tab:performancecomparison}
\begin{tabular}{|l|l|r|r|r|}
\hline
\textbf{Dataset} & \textbf{Model}  & \textbf{Acc.} & \textbf{F-1} & \textbf{Avg. CS} \\ \hline
    & CNN              & \textbf{0.83}     & \textbf{0.83}      & 0.79     \\ \cline{2-5} 
    & GPT-4o                & 0.62              & 0.54               & \textbf{0.93}        \\ \cline{2-5} 
\multirow{-3}{*}{\textbf{Chest X-ray}}    & Llama3.2-vision       & 0.65        & 0.64  & 0.92     \\ \hline
& CNN              & \textbf{0.98}     & \textbf{0.99}      & \textbf{0.99}             \\ \cline{2-5} 
 & GPT-4o                  & 0.60              & 0.58               & 0.93                \\ \cline{2-5} 

\multirow{-3}{*}{\textbf{MRI}} & Llama3.2-vision         & 0.52              & 0.51               & 0.85                                \\ \hline
 & CNN            & \textbf{0.91}     & \textbf{0.90}      & \textbf{0.94}                       \\ \cline{2-5} 
 & GPT-4o                 & 0.22              & 0.14               & 0.91                                \\ \cline{2-5} 
\multirow{-3}{*}{\textbf{Chest CT}}               & Llama3.2-vision        & 0.50              & 0.48               & 0.80                                \\ \hline

\end{tabular}
\vspace{-0.1 in}
\end{table}

\begin{figure*}[t]
    \centering
    \subfigure[Average Execution Time]{\label{fig:time}
        \includegraphics[width=0.325\textwidth]{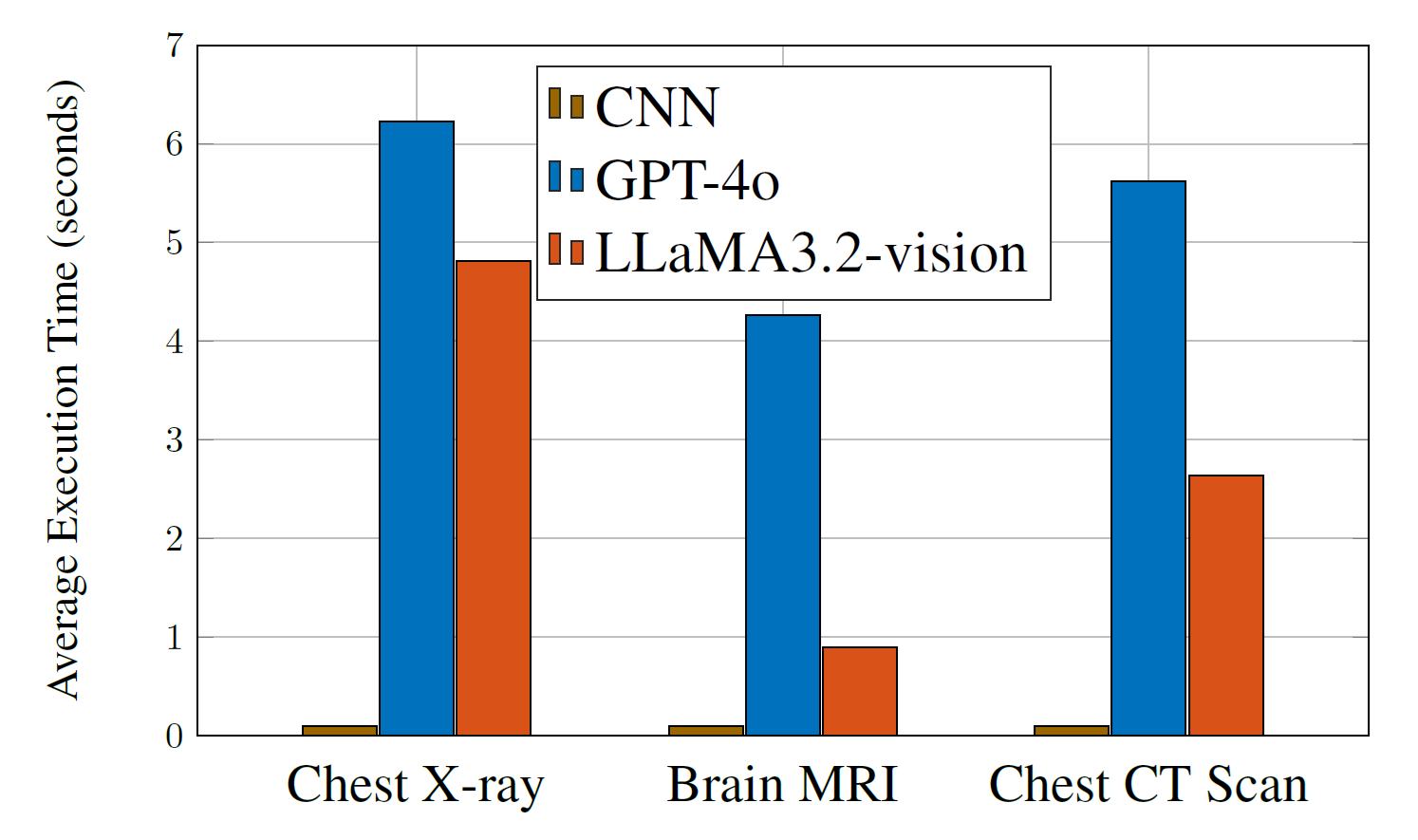}}
    \subfigure[Average Energy Consumption]{\label{fig:energy}
        \includegraphics[width=0.31\textwidth]{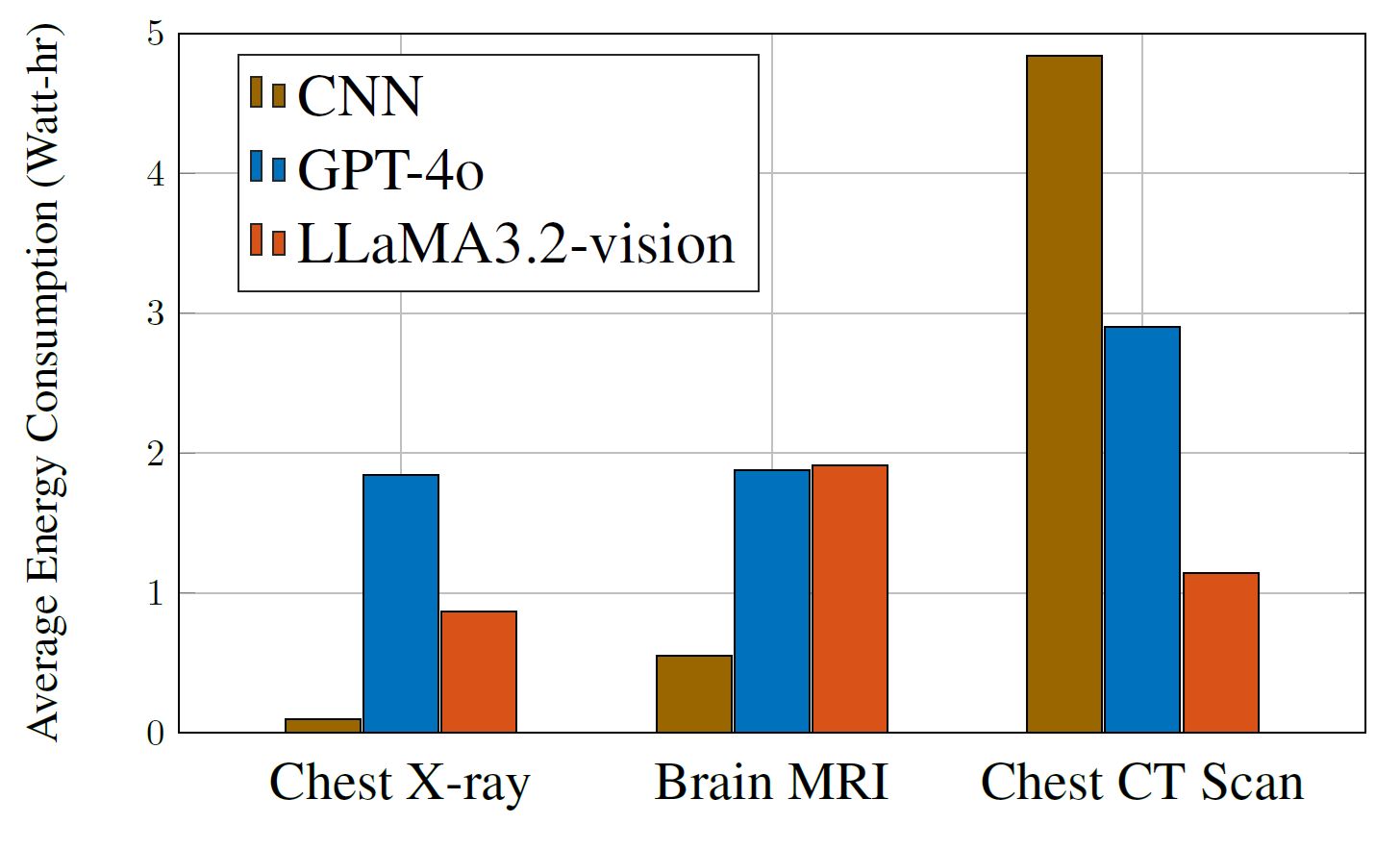}}
    \subfigure[Average $CO_2$ Emission]{\label{fig:co2}
        \includegraphics[width=0.33\textwidth]{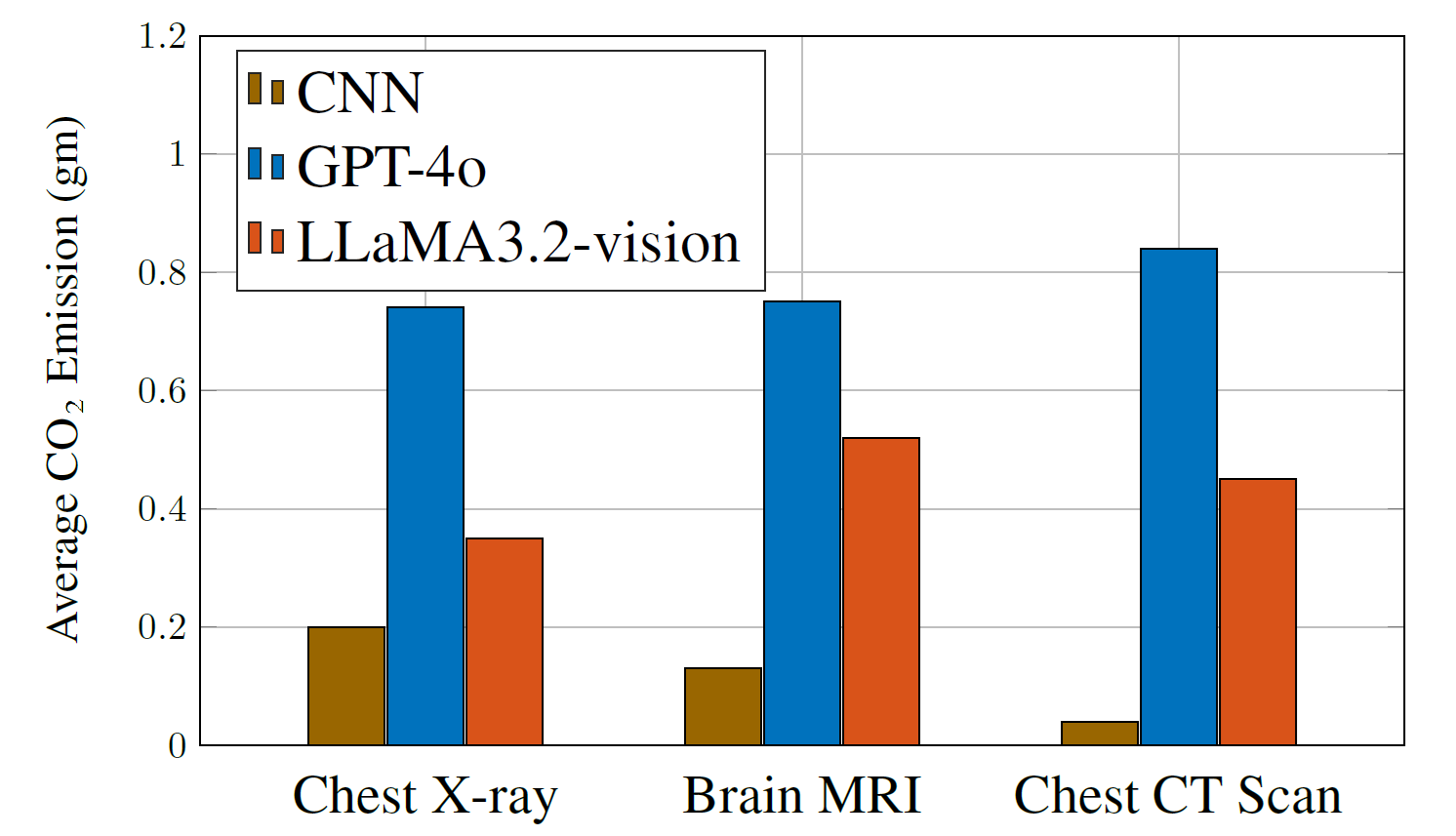}}
        \vspace{-0.1 in}
    \caption{\small Comparative performance of CNNs and LLMs across three medical imaging datasets: chest X-ray, MRI, and chest CT scans}
    \label{fig:chart}
        \vspace{-0.1 in}
\end{figure*}

\subsection{Results Overview}
Table~\ref{tab:performancecomparison} summarizes comparative performance of CNNs and LLMs (GPT-4o and Llama3.2-vision) across medical imaging datasets: chest X-ray, Brain MRI, and chest CT scans. The results indicate notable variations in accuracy, F1-score, and confidence, revealing each model's strengths and limitations.

\vspace{0.05 in}
\noindent \textbf{Chest X-ray:}
The CNN achieves the highest overall performance, with precision, recall, accuracy, and F1-score of 0.83, surpassing both GPT-4o and Llama3.2-vision. Its average confidence score of 0.79 indicates reliable prediction confidence. 
GPT-4o, while achieving the highest confidence score (0.93), has a relatively low accuracy (0.62) and F1-score (0.54), reflecting overconfidence in its predictions. Llama3.2-vision shows slightly better accuracy (0.65) and F1-score (0.64) than GPT-4o, but it still underperforms compared to the CNN.

\vspace{0.05 in}
\noindent \textbf{Brain MRI:}
CNNs demonstrate exceptional performance on MRI datasets, with near-perfect precision, recall, accuracy, and F1-score (all above 0.98). This underscores their capability to handle well-defined imaging patterns. In contrast, GPT-4o and Llama3.2-vision exhibit poor accuracy and F1-scores (below 0.60), suggesting limited effectiveness. 
Despite a high confidence score (0.93), GPT-4o exhibits a mismatch between confidence and actual performance.

\vspace{0.05 in}
\noindent \textbf{Chest CT Scan:}
For chest CT scans, CNNs again lead in performance, achieving an accuracy and F1-score of around 0.91. GPT-4o performs poorly with an accuracy of 0.22 and an F1-score of 0.14, despite maintaining a confidence score of 0.91. Llama3.2-vision performs slightly better, with an accuracy of 0.50 and an F1-score of 0.48, but both LLMs lag significantly behind CNNs for this modality.

\vspace{-0.1 in}
\subsection{Computational Efficiency Analysis}
The Fig.~\ref{fig:chart} compares CNNs, GPT-4o, and Llama3.2-vision across three datasets (Chest X-ray, MRI, and Chest CT scan) in terms of execution time, energy consumption, and CO\textsubscript{2} emissions. We calculated energy consumption and CO\textsubscript{2} emissions~\cite{c02} from execution time, power usage, and average carbon intensity. CNNs demonstrate the best computational efficiency, with minimal execution time, energy usage, and environmental impact, making them highly suitable for real-time medical diagnostics. In contrast, LLMs (GPT-4o and Llama3.2-vision) exhibit significantly higher execution times and energy consumption, particularly Llama3.2-vision, which incurs higher costs across all metrics. CNN inference times for chest CT scans can be longer due to complexity of volumetric images. These results illustrate the trade-offs between CNNs' computational efficiency and LLMs' multimodal reasoning, indicating that CNNs are better for scalable clinical applications, while LLMs excel in complex contextual tasks.

\vspace{-0.1 in}

\subsection{Effect of Enhanced Data Filtering}

The results of our enhanced data filtering process, shown in Table \ref{tab:filtcomp}, highlight its effectiveness with the chest X-ray dataset. Using GPT-4o alone yields 62\% accuracy, but adding a filtering step (Sec.~\ref{sec:filtering}) boosts accuracy to 82\% with an average confidence score of 0.93. This filtering also cuts execution time from 6.23  seconds to 2.35  seconds and reduces energy consumption from 1.84 watt-hours to 1.65 watt-hours. We anticipate similar gains with other LLMs across various imaging modalities, but results may vary by dataset complexity.

\begin{table}[t]
 \centering
 \caption{\small Comparison of performance with and without data filtering. The symbol \arrowUpDarkGreen\ indicates a desirable increase in the metric, whereas \arrowDownDarkGreen\ indicates a desirable decrease.} 
 \vspace{0.1 in}
 \label{tab:filtcomp}

 \begin{tabular}{|c|c|c|}
 \hline
 \textbf{Metric} & \textbf{w/o filtering} & \textbf{with filtering} \\
 \hline
 Accuracy & 62\% & 82.01\% \arrowUpDarkGreen \\
 \hline
 Avg. Confid. Score & 0.93 & 0.93 {\color{blue}=} \\
 \hline
 Avg. Execut. Time & 6.23 s & 2.35 s \arrowDownDarkGreen \\
 \hline
 Avg. Energy Consump. & 1.84 W-H & 1.65 W-H \arrowDownDarkGreen \\
 \hline

 \end{tabular}
  \vspace{-0.1 in}
 \end{table}

 \vspace{-0.1 in}
\subsection{Additional Insights}

\noindent \textbf{Limitations of traditional LLMs:}
While the LLMs, including GPT-4o and Llama3.2-vision, show moderate success in handling medical image classification tasks, their confidence scores, particularly for GPT-4o, are disproportionately high relative to accuracy, highlighting need for improved calibration. The lower performance of LLMs suggests that they require further adaptation or fine-tuning for image-specific tasks.

\noindent \textbf{Strengths of Data Filtering in LLM}
Providing necessary contextual details to the Large Language Model (LLM), a framework can simplify the decision-making cycle,  reduce processing times and energy usage, thereby boosting overall efficiency and lower environmental impact. Such a streamlined approach fosters more robust and eco-friendly medical image classification while maintaining high accuracy.

\vspace{0.07 in}
 \noindent \textbf{Insights on Model Calibration:}
The calibration of confidence scores is a critical aspect of deploying AI systems in clinical settings. CNNs demonstrate better alignment between confidence and accuracy, making them more suitable for high-stakes applications. LLMs need calibration to enhance trust in clinical workflows.

\vspace{0.07 in}
 \noindent \textbf{Dataset Complexity:}
 The performance differences across datasets reveal that MRI datasets achieve higher accuracy due to distinct and consistent patterns, while chest CT scans are more challenging because of complex and subtle features, highlighting each model's strengths and limitations.

%% file: conclusion2.tex
\section{Conclusion and Future Work}
\label{conclusion}

This study compared CNNs and LLMs for medical image classification, highlighting their strengths and limitations. CNNs demonstrated superior accuracy and efficiency, making them well-suited for real-time clinical applications, while LLMs showed potential in multimodal reasoning but suffered from high computational costs. Our findings emphasize the trade-offs between accuracy, reliability, and resource consumption, offering practical insights for AI deployment in healthcare.
Future works will focus on hybrid models that merge deep neural networks with LLMs, enhancing medical diagnostics through multimodal AI~\cite{chexzero}. This integration aims to improve both accuracy and interpretability by combining visual features with contextual reasoning.
Additionally, our research offers a benchmark with LLM-predicted labels and detailed medical image analysis, paving the way for new AI research in medical imaging. Furthermore, explainable AI techniques can enhance transparency and trust among clinicians, aiding the real-world adoption of these technologies in healthcare.
\vspace{-0.1 in}